\documentclass[12pt]{article}
\usepackage[]{epsfig}
\usepackage[centertags]{amsmath}
\usepackage{amsfonts}
\usepackage{amssymb}
\usepackage[english]{babel}
\usepackage{amsmath, amsthm, slashed,url}

\begin{document}

\title{\bf Self-Dual Soliton Solution in a Generalized Jackiw-Pi Model  }
\author{
Lucas Sourrouille
\\
{\normalsize \it Universidad Nacional Arturo Jauretche
}\\ {\normalsize\it Florencio Varela, Buenos Aires, Argentina}
\\
{\footnotesize  sourrou@df.uba.ar} } \maketitle

\abstract{We consider a generalization of Jackiw-Pi model by introducing a nonstandard kinetic term. We present a Bogomolnyi framework for this theory and as a particular case we show that the Bogomolnyi equations of Chern-Simons Higgs theory can be obtained. Finally, the dimensionally reduced theory is analyzed and novel solitonic equations emerge. }

\vspace{0.5cm}



\vspace{1cm}
\section{Introduction}
The two dimensional matter field interacting with gauge fields whose dynamics is governed by a Chern-Simons term support soliton solutions\cite{Jk}, \cite{hor}. These models have the particularity to became auto-dual when the self-interactions are suitably chosen\cite{JW,JP}. When this occur the model presents particular mathematical and physics properties, such as the supersymmetric extension of the model\cite{LLW}, and the reduction of the motion equation to first order derivative equation\cite{Bogo}. The Chern-Simons gauge field inherits its dynamics from the matter fields to which it is coupled, so it may be either relativistic\cite{JW} or non-relativistic\cite{JP}. In addition the soliton solutions are of topological and non-topological nature\cite{JLW}.

In the recent years, theories with nonstandard kinetic term, named $k$-field models, have received much attention. The $k$-field models are mainly in connection with effective cosmological models\cite{BMV,APDM} as well as the strong interaction physics, strong gravitational waves\cite{MV} and dark matter\cite{APL}. One interesting aspect to analyze in these models concern to its topological structure. In this context several studies have been conducted, showing that the $k$-theories can support topological soliton solutions both in models of matter as in gauged models\cite{BAi,SG}. These solitons have certain features such as their characteristic size, which are not necessarily those of the standard models\cite{B}. Other interesting aspects are that they do not interact at large distances and they are, in general, not self-dual.

In this paper we are interested in studying the Jackiw-Pi model with a generalized dynamics. The so call Jackiw-Pi model is a nonrelativistic and Galilean invariant model which is also self-dual\cite{JP}. Here, we will show that introducing a nonstandard dynamics in the Jackiw-Pi Lagrangian, via a function $\omega$ depending on the Higgs field, we can obtain self-dual or Bogomolnyi equations by minimizing the energy functional of the model. As particular case, we will show that choosing a suitable $\omega$, the Bologmolnyi equations of Chern-Simons Higgs theory can be obtained. Finally, we will study the the dimensional reduction of the model to $(1+1)$-dimensions and we will arrive to the existence of interesting soliton solutions in the system.

\section{The Model}
Let us start by considering the model proposed by Jackiw and Pi\cite{JP}
\begin{eqnarray}
S = \int \,\, d^3 x & \Big( \frac{\kappa}{2}\epsilon^{\mu \nu \rho} A_\mu \partial_\nu A_\rho
+i\phi^* D_0 \phi - \frac{1}{2m} |D_i \phi|^2 + \lambda|\phi|^4 \Big)
\label{Ac22}
\end{eqnarray}
This is a nonrelativistic model where the gauge fields dynamics is dictated by a Chern-Simons term and matter is represented by scalar field $\phi(x)$. The covariant derivative is defined as  $D_{\mu}= \partial_{\mu} + ieA_{\mu}$ $(\mu =0,1,2)$. The metric tensor is  $g^{\mu \nu}=(1,-1,-1)$ and $\epsilon^{\mu\nu\lambda}$ is the totally antisymmetric tensor such that $\epsilon^{012}=1$.

The field equations corresponding to this action are

\begin{eqnarray}
i D_0 \phi = - \frac{1}{2m} D_i^2 \phi- 2\lambda|\phi|^2\phi
\nonumber \\[3mm]
B=\frac{e}{\kappa} \rho
\nonumber \\[3mm]
E^i = -\frac{1}{\kappa} \epsilon^{i j} j_i
\label{EqM4}
\end{eqnarray}
where $\rho= |\phi|^2$ and $j^i = -\frac{i}{2m}\Big(\phi^* D^i \phi - (D^i \phi)^* \phi \Big)$. The second of this equations is the Chern-Simons Gauss law, which can be integrated, over the entire plane, to obtain the important consequence that any object with charge $Q =e\int d^2 x \rho$ also carries magnetic flux $\Phi = \int B d^2 x$ \cite{Echarge}:

\begin{eqnarray}
\Phi =\frac{1}{\kappa} Q
\end{eqnarray}

Here, we are interested in time-independent soliton solutions that ensure the finiteness of the action (\ref{Ac22}). These are the stationary points of the energy which for the static field configuration reads

\begin{eqnarray}
E = \int \,\, d^2 x & \Big(  \frac{1}{2m} |D_i \phi|^2 - \lambda|\phi|^4 \Big)
\label{EJP}
\end{eqnarray}
In order to find the minimum of the energy, the expression (\ref{EJP}) can be rewritten as

\begin{eqnarray}
E = \int \,\, d^3 x & \Big(  \frac{1}{2m} |D_{\pm} \phi|^2 + (-\lambda \mp \frac{e^2}{2m\kappa})|\phi|^4 \Big)
\label{}
\end{eqnarray}
where we have used the Chern-Simons Gauss law and the identity
\begin{eqnarray}
|D_i \phi|^2 = |( D_1 \pm iD_2)\phi|^2 \mp eB|\phi|^2 \pm m\epsilon^{ij} \partial_i J_j
\label{iden}
\end{eqnarray}
Thus, with the self-dual coupling
\begin{eqnarray}
\lambda= \mp \frac{e^2}{2m\kappa}
\end{eqnarray}
and sufficiently well behaved fields so that the integral over all space of $\epsilon^{ij} \partial_i J_j$ vanishes, the energy becomes
\begin{eqnarray}
E = \int \,\, d^3 x &  \frac{1}{2m} |D_{\pm} \phi|^2
\label{}
\end{eqnarray}
Thus, the energy is bounded below by zero, and this lower bound is saturated by fields obeying the first order self-duality equations
\begin{eqnarray}
(D_1 \pm iD_2)\phi =0
\nonumber \\[3mm]
B = \frac{e}{\kappa} \rho
\label{}
\end{eqnarray}

Following the same idea of the works cited in Ref.\cite{BAi}, we will consider, here, a generalization of the Jackiw-Pi model described by the action

\begin{eqnarray}
S = \int \,\, d^3 x & \Big( \frac{\kappa}{2}\epsilon^{\mu \nu \rho} A_\mu \partial_\nu A_\rho
+i\omega(\rho)\phi^* D_0 \phi - \frac{1}{2m} |D_i \phi|^2 - V(\rho) \Big)
\label{Ac3}
\end{eqnarray}
where we have replaced the usual kinetic term $i\phi^* D_0 \phi$ by a more generalized term $i\omega(\rho)\phi^* D_0 \phi$. Here, $\omega(\rho)$ is, in principle, an arbitrary function of the complex scalar field $\phi$ and $V(\rho)$ is the scalar field potential to be determined below.

The equations of motion for this system are given by
\begin{eqnarray}
i\Big(\frac{\partial\omega(\rho)}{\partial \phi^*}\phi^*D_0 \phi + \omega(\rho)D_0 \phi\Big) = - \frac{1}{2m} D_i^2 \phi + \frac{\partial V(\rho)}{\partial \phi^*}
\nonumber \\[3mm]
B=\frac{e}{\kappa} \omega(\rho)\rho
\nonumber \\[3mm]
E^i = -\frac{1}{\kappa} \epsilon^{i j} j_i
\label{EqM3}
\end{eqnarray}
where the two first equations differ form those present in Eq.(\ref{EqM4}) by the presence of the function $\omega(\rho)$.

The theory may be descried in terms of the Hamiltonian formulation as
\begin{eqnarray}
H = \int \,\, d^2 x & \Big(  \frac{1}{2m} |D_i \phi|^2 + V(\rho) \Big)
\label{}
\end{eqnarray}
which may be rewritten using the Gauss Law and the identity (\ref{iden}) in the form
\begin{eqnarray}
E = \int \,\, d^3 x & \Big(  \frac{1}{2m} |D_{\pm} \phi|^2 \mp \frac{e^2}{2m\kappa} \omega(\rho)\rho^2 + V(\rho) \Big)
\label{E2x}
\end{eqnarray}
In order to relate the solutions in this theory with those present in the Chern-Simons Higgs theory, we may choose, as a particular case, the following $\omega(\rho)$ function
\begin{eqnarray}
\omega(\rho) = 2m\frac{e^2}{\kappa} (\rho -1)
\label{omega}
\end{eqnarray}
Then, the energy functional (\ref{E2x}) is written as
\begin{eqnarray}
E = \int \,\, d^3 x & \Big(  \frac{1}{2m} |D_{\pm} \phi|^2 \mp \frac{e^4}{\kappa^2} \rho^2(\rho -1) + V(\rho) \Big)
\label{E11}
\end{eqnarray}
and the Gauss law of the equation (\ref{EqM3}) takes the form
\begin{eqnarray}
B=2m\frac{e^3}{\kappa^2} (\rho -1)\rho
\end{eqnarray}
The form of the potential $V(\rho)$ that we choose is motivated by the desire to find self-dual soliton solution.
Thus, if we choose the potential as
\begin{eqnarray}
V(\rho) = \pm \frac{e^4}{\kappa^2} \rho(\rho-1)^2
\label{pot}
\end{eqnarray}
and replace it in the expression (\ref{E11}) we arrive to following expression of energy functional
\begin{eqnarray}
E = \int \,\, d^3 x & \Big(  \frac{1}{2m} |D_{\pm} \phi|^2 \mp \frac{e}{2m} B \Big)
\label{Ener}
\end{eqnarray}
which is bounded below by a multiple of the magnitude
of the magnetic flux (for positive flux we choose the lower signs, and for negative flux we choose
the upper signs):
\begin{eqnarray}
E \geq \frac{e}{2m} |\Phi|
\end{eqnarray}
Here, the magnetic flux is determined by the requirement of finite energy. This implies that the covariant derivative must vanish asymptotically, which fixes the  behavior of the gauge field $A_i$. Then we have
\begin{eqnarray}
\Phi = \int \,\,d^2 x B = \oint_{|x|=\infty} \,\, A_i dx^i  = 2\pi N
\end{eqnarray}
where $N$ is a topological invariant which takes only integer values. It is interesting to remark, here, the existence of the topological bound which is not present in the Jackiw-Pi model. So, this a non-relativistic model with a topological bound and therefore we shall expect to find topological solitons.

This bound is saturated by fields satisfying the first-order self-duality equations
\begin{eqnarray}
& &D_{\pm}\phi = ( D_1 \pm iD_2)\phi =0
\\
& &
B=2m\frac{e^3}{\kappa^2} (\rho -1)\rho
\end{eqnarray}

These equations may be compared with the self-duality equations of the Chern-Simons Higgs theory. We can note that if we fix $m=1$ and choose the plus sign in the potential expression (\ref{pot}) we arrive to the Chern-Simons Higgs self-duality equations
\begin{eqnarray}
& &( D_1 + iD_2)\phi =0
\\
& &
B=2m\frac{e^3}{\kappa^2} (\rho -1)\rho
\end{eqnarray}
On the other hand the anti-self-duality equations may be obtained by choosing the function $\omega(\rho)$ as
\begin{eqnarray}
\omega(\rho) = -2m\frac{e^2}{\kappa} (\rho -1)
\end{eqnarray}
In that case, if we desire to arrive to the expression (\ref{Ener}), we must choose the following potential term
\begin{eqnarray}
V(\rho) = \mp \frac{e^4}{\kappa^2} \rho(\rho-1)^2
\label{pot2}
\end{eqnarray}
Then, the Bogomolnyi equations becomes
\begin{eqnarray}
& &D_{\pm}\phi = ( D_1 \pm iD_2)\phi =0
\\
& &
B= -2\frac{e^3}{\kappa^2} (\rho -1)\rho
\end{eqnarray}
Choosing $V(\rho) = \frac{e^4}{\kappa^2} \rho(\rho-1)^2$ we obtain the anti-self-duality equations
\begin{eqnarray}
& &D_{-}\phi = ( D_1 -iD_2)\phi =0
\\
& &
B= -2\frac{e^3}{\kappa^2} (\rho -1)\rho
\end{eqnarray}
Thus, we can obtain the same Bogomolnyi equations as those present in the Chern-Simons Higgs model. The difference lies in the fact that in our case, we have dealing with a non-relativistic model and we have imposed the constraint $m=1$.
Another interesting fact is that, here, we expect to find both topological and nontopological soliton solutions, just as in Chern-Simons Higgs theory. Also it is worth noting that the Bogonolnyi equations in Chern-Simons Higgs theory are neither solvable nor integrable. However, numerical solutions can be found using a radial vortex-like ansatz\cite{JW,JLW}.

An important comment is that the generalized Jackiw-Pi model, studied here, is a self-dual model. This is important because the generalized pure Chern-Simons system previously explored are not self-dual (For instance see Ref.\cite{BAi}).

\section{Dimensional reduction and the solitonic solution}
In this section we are interested in analyzing the dimensional reduction of the model (\ref{Ac3}) as well as in studying the soliton solution in the dimensionally reduced model. In order to analyze the lineal problem\cite{my,JPR}, it is natural to consider a dimensional reduction of the action (\ref{Ac3}) by suppressing dependence on the second spacial coordinate, renaming $A_y$ as $B$. Then, the action (\ref{Ac3}) becomes
\begin{eqnarray}
S = \int \,\, d^2 x &\Big(& \kappa ( A_0 \partial_x B + B\partial_0 A_1)
+i\omega(\rho)\phi^* D_0 \phi \nonumber \\
&-& \frac{1}{2m} |D_x \phi|^2 -\frac{e^2}{2m} B^2\rho - V(\rho) \Big)
\label{Ac4}
\end{eqnarray}
Notice that the Gauss law constraint for this action is
\begin{eqnarray}
\partial_x B=\frac{e}{\kappa} \omega(\rho)\rho\;,
\label{gauss1}
\end{eqnarray}
which can be solved as
\begin{eqnarray} B(x)=\frac{e}{2 \kappa} \int dz \epsilon(x-z) \omega(\rho(z))\rho(z)
\label{gauss2} \end{eqnarray}
where $\epsilon(x)=\theta(x)-\theta(-x)$ is the odd step function.

By using the  Gauss law and the  explicit form of $B$ given by equation (\ref{gauss2}), the action can be written simply as
\begin{eqnarray}
S = \int \,\, d^2 x &\Big[& \Big(\frac{e}{2 } \int dz \epsilon(x-z) \omega(\rho(z))\rho(z)\Big)\partial_0 A_1)
+i\omega(\rho)\phi^* \partial_0 \phi \nonumber \\
&-& \frac{1}{2m} |D_x \phi|^2 -\frac{e^2}{2m} B^2\rho - V(\rho) \Big]
\label{Ac5}
\end{eqnarray}
Following the  Ref.\cite{JPR}, the gauge field $A_x$ may be eliminate from  the action (\ref{Ac5}) by a gauge transformation.
Indeed, after transforming the matter field as

\begin{eqnarray}
\phi(x)\rightarrow e^{-i\alpha(x)} \phi(x)
\end{eqnarray}
with
\begin{eqnarray}
\alpha(x)=\frac{e}{2}\int dz \epsilon(x-z) A_x(z)
\label{alpha}
\end{eqnarray}
we arrive to the following action
\begin{eqnarray}
S = \int \,\, d^2 x &\Big( i\omega(\rho)\phi^* \partial_0 \phi
- \frac{1}{2m} |\partial_x \phi|^2 -\frac{e^2}{2m} B^2\rho - V(\rho) \Big)
\label{Ac6}
\end{eqnarray}
Consider, now, the derivation of the Bogomolnyi equations in the reduced model (\ref{Ac6}). As discussed in Ref.\cite{OH,my1} the field $B$ plays an important role in the derivation of the self-dual equations. Indeed the expression (\ref{gauss2}) of $B$,  involves the existence of a novel soliton solution.
Using the relation
\begin{eqnarray}
\int \,\, d^2 x \Big(|( \partial_x + \gamma e B)\phi|^2 + \gamma e\partial_x B \rho \Big) = \int \,\, d^2 x\Big(|\partial_x \phi|^2 + e^2B^2\rho \Big)
\end{eqnarray}
we can rewrite the action (\ref{Ac6}) as
\begin{eqnarray}
S = \int \,\, d^2 x &\Big( i\omega(\rho)\phi^* \partial_0 \phi
- \frac{1}{2m} |( \partial_x + \gamma eB)\phi|^2 -\frac{\gamma e}{2m} \partial_x B \rho - V(\rho) \Big)
\label{Ac7}
\end{eqnarray}
Here, $\gamma =\pm 1$. The Gauss law (\ref{gauss1}) may be used to replace the derivative of the field $B$ in the action. Then we have, in the static field configuration, that the Hamiltonian associated to action is written as
\begin{eqnarray}
H= \int \,\, d x &\Big(
\frac{1}{2m} |( \partial_x + \gamma eB)\phi|^2 +\frac{e^2\gamma }{2m\kappa} \omega(\rho) \rho^2 + V(\rho) \Big)
\label{H}
\end{eqnarray}
As in the $(2+1)$-dimensional case, we can choose $\omega(\rho) = 2m\frac{e^2}{\kappa} (\rho -1)$ and $V(\rho) = \mp \frac{e^4}{\kappa^2} \rho(\rho-1)^2$ to obtain
\begin{eqnarray}
H= \int \,\, d x &\Big(
\frac{1}{2m} |( \partial_x + \gamma eB)\phi|^2 +\frac{e\gamma }{2m} \partial_x B \Big)
\label{H1}
\end{eqnarray}
Since the field $B$ must be zero in the boundary the last term in the Hamiltonian vanish and we have
\begin{eqnarray}
H= \int \,\, d x &
\frac{1}{2m} |( \partial_x + \gamma eB)\phi|^2
\label{H2}
\end{eqnarray}
This is non-negative and therefore takes the minimum when the $\phi$ satisfies
\begin{eqnarray}
& &(\partial_x +\gamma
B)\phi=0
\end{eqnarray}
We can write this equation in a more explicitly form by using the Eq.(\ref{gauss2})
\begin{eqnarray}
\partial_x\phi(x) +
\frac{\gamma e}{2 \kappa} \int dz \epsilon(x-z) \omega(\rho(z))\rho(z)\phi(x)= 0
\label{Eq1}
\end{eqnarray}
To solve (\ref{Eq1}), we can proceed as in Ref.\cite{my1}. Thus, we assume that $\phi$ may be written as $\phi=\sqrt{\rho}$, which leads to
\begin{eqnarray}
\frac{1}{2}\partial_x (\log \rho(x)) + \frac{e\gamma}{2 \kappa} \int dz \epsilon(x-z)\omega(\rho(z))\rho(z) =0
\end{eqnarray}
Differentiating the above equation with respect to x, we arrive to the following 1-dimensional Liouville type equation
\begin{eqnarray}
\frac{1}{2}\partial_x^2 (\log \rho(x)) + \frac{e\gamma}{\kappa} \omega(\rho(x))\rho(x) =0
\end{eqnarray}
This is the general Bogomolnyi equation corresponding to dimensionally reduced model. In particular we are analyzing the case $\omega(\rho) = 2m\frac{e^2}{\kappa} (\rho -1)$, so that the previous equation becomes
\begin{eqnarray}
\frac{1}{2}\partial_x^2 (\log \rho(x)) + \frac{2e^3\gamma}{\kappa^2} (\rho -1)\rho(x) =0
\label{Lio}
\end{eqnarray}
This equation presents two types of solution, one derived from the topological solution and the other from nontopological solution present in the two dimensional Chern-Simons Higgs theory. Let us start by considering the solution derived from the topological case. For this case we propose as a solution the following series

\begin{eqnarray}
\rho = 1 + \sum_{n=1}^\infty a_n \rm{sech}^n (b x)
\label{serie}
\end{eqnarray}
where $a_n$ are the real coefficients of series and $b$ is a real constant. To check that this is really a solution, we rewrite Eq.(\ref{Lio} ) as
\begin{eqnarray}
-(\partial_x \rho)^2 + (\partial_x^2 \rho)\rho + \upsilon \rho^3(\rho -1) =0
\label{Lio2}
\end{eqnarray}
with $\upsilon = \frac{4e^3\gamma}{ \kappa^2}$. Inserting the series (\ref{serie}) into equation (\ref{Lio2}), we obtain
\begin{eqnarray}
& &\sum_{n,m=1}^\infty \Big[ -a_n a_m m n b^2 + n^2a_n a_m b^2 + 3\upsilon a_n a_m \Big]  \rm{sech}^{n+m}(bx) +
\nonumber \\
& &\sum_{n,m=1}^\infty \Big[ a_n a_m n m b^2 - n^2a_n a_m b^2 - n a_n a_m b^2 \Big] \rm{sech}^{n+m+2} (bx) -
\nonumber \\
& &\sum_{n=1}^\infty (n^2 +n)a_n b^2 \rm{sech}^{2+n} (bx) + \sum_{n=1}^\infty \Big[ n^2 a_n b^2 + \upsilon a_n \Big] \rm{sech}^n (bx) +
\nonumber \\
& &3\upsilon \sum_{n,m,i=1}^\infty a_n a_m a_i \rm{sech}^{n+m+i}(bx) + \upsilon \sum_{n,m,i,j=1}^\infty a_n a_m a_i a_j \rm{sech}^{n+m+i+j}(bx)=0
\label{exp}
\end{eqnarray}
where, in this last equation, we have used the relation
\begin{eqnarray}
\tanh^2 (bx) =1 - \rm{sech}^2 (bx)
\end{eqnarray}
Since the expression (\ref{exp}) is an expansion of powers of $\rm{sech}(bx)$ each coefficient of different powers must vanish separately. This implies that for the coefficient of $\rm{sech}(bx)$, we have following relation
\begin{eqnarray}
b^2=-\upsilon
\end{eqnarray}
whereas that for the coefficients of $\rm{sech}^2(bx)$ and $\rm{sech}^3(bx)$, we deduce

\begin{eqnarray}
a_2= 2a_1^2
\end{eqnarray}
and

\begin{eqnarray}
a_3 = \frac{13}{8}a_1^3 -\frac{1}{2}a_1
\end{eqnarray}
This procedure may be repited in order to determine the successive coefficients.

There is another type of solution, which satisfies the boundary condition $\rho\to 0$ as $|x|\to \infty$. This case is more complicated and we are not able to propose an analytic expression for the solution. However, it is not difficult to obtain approximate asymptotic solution for large $|x|$. In this limit the equation $(\ref{Lio})$ may be approximated by
\begin{eqnarray}
\partial_x^2 (\log \rho(x)) + \upsilon (\rho -1)\rho(x) =0
\label{}
\end{eqnarray}
As in the previous case the solution exists for $\upsilon<0$;
\begin{eqnarray}
\rho= a\rm{sech}^2(bx)\;,
\end{eqnarray}
where the $a$ and $b$ are related by $b^2= -\frac{a \upsilon}{2}$. Thus if the solution exist, should approach to zero exponentially.   

\section{Conclusion}

We have studied a generalized Jackiw-Pi model by introducing a non-standard dynamics $\omega(\rho)$ in the original Jackiw-Pi Lagrangian. It was shown that this model support Bogomolnyi equations and soliton solutions therein, which represent an important fact because the other soliton solutions found in generalized pure Chern-Simons models are not self-dual. In particular have shown that choosing a suitable function $\omega(\rho)$  and a sixth-order self-dual potential, the energy of the model is bounded below by a topological number. In addition, the resulting Bogomolnyi equations are those of the Chern-Simons Higgs theory.

The introduction of the function $\omega(\rho)$ has also consequences at quantum level. Although the model may be viewed as the second quantized $N$-particle system interacting with a Chern-Simons gauge field, its dynamics is modified by $\omega(\rho)$, and therefore differs from dynamics of the Jackiw-Pi model. In this case the quantum field equation is 
\begin{eqnarray}
i\partial_t \Phi(x)=[\Phi(x), H] 
\end{eqnarray}
and the gauge field should be subject to the constraint 
\begin{eqnarray}
B=\frac{e}{\kappa} \omega(\rho)\rho 
\end{eqnarray}

Finally, it is interesting to mention that the $(1+1)$-dimensional model, obtained by dimensional reduction of the generalized Jackiw-Pi, presents novel solitons solutions. One type of these solitons has a topological origin and have been able to find the analytical expression, the other has non-topological origin and we have found its asymptotic behavior.

\end{document}